\documentclass[conference,a4paper]{IEEEtran}

\usepackage[noadjust]{cite}

\ifCLASSINFOpdf
\else
\usepackage[dvips]{graphicx}
\DeclareGraphicsExtensions{.eps}
\fi

\usepackage{psfrag}

\usepackage[cmex10]{amsmath}
\usepackage{amssymb}
\def\uv{\mathbf{u}}
\def\cv{\mathbf{c}}
\def\xv{\mathbf{x}}
\def\hv{\mathbf{h}}
\def\yv{\mathbf{y}}
\def\zv{\mathbf{z}}
\def\wv{\mathbf{w}}
\def\muv{\boldsymbol{\mu}}
\def\Sgm{\boldsymbol{\Sigma}}
\def\I{\mathcal{I}}
\def\A{\mathcal{A}}
\def\N{\mathcal{N}}
\def\P{\mathcal{P}}
\def\D{\mathcal{D}}

\def\S{\mathcal{S}}
\def\M{\mathcal{M}}

\newtheorem{algorithm}{Algorithm}

\begin{document}
%
\title{Distributed Iterative Processing for Interference Channels with Receiver Cooperation}

\author{
\IEEEauthorblockN{
Mihai-Alin Badiu\IEEEauthorrefmark{1}\IEEEauthorrefmark{2},
Carles Navarro Manch\'on\IEEEauthorrefmark{1},
Vasile Bota\IEEEauthorrefmark{2} and
Bernard Henri Fleury\IEEEauthorrefmark{1}}
\IEEEauthorblockA{
\IEEEauthorrefmark{1}
Aalborg University, Denmark
}
\IEEEauthorblockA{
\IEEEauthorrefmark{2}
Technical University of Cluj-Napoca, Romania
}
}

\maketitle

\begin{abstract}
We propose a framework for the derivation and evaluation of distributed iterative algorithms for receiver cooperation in interference-limited wireless systems.
Our approach views the processing within and collaboration between receivers as the solution to an inference problem in the probabilistic model of the whole system. The probabilistic model is formulated to explicitly incorporate the receivers' ability to share information of a predefined type.
We employ a recently proposed unified message-passing tool to infer the variables of interest in the factor graph representation of the probabilistic model. 
The exchange of information between receivers arises in the form of passing messages along some specific edges of the factor graph; the rate of updating and passing these messages determines the communication overhead associated with cooperation.
Simulation results illustrate the high performance of the proposed algorithm even with a low number of message exchanges between receivers.
\end{abstract}



%
\IEEEpeerreviewmaketitle

\section{Introduction}
Cooperation in interference-limited wireless networks has the potential to significantly improve the system performance~\cite{Gesbert10}. Additionally, variational techniques for Bayesian inference~\cite{Wainwright2008} are proven extremely useful for the design of iterative receiver architectures in non-cooperative scenarios. Hence, using such inference methods to design iterative algorithms for receiver cooperation could be beneficial.
%
%
%
%

Algorithms based on belief propagation (BP) are proposed in~\cite{Grant04,Aktas08} for distributed decoding in the uplink of cellular networks with base-station cooperation, assuming simple network models, uncoded transmissions and perfect channel knowledge at the receivers; it is shown that the performance of optimal joint decoding can be achieved with decentralized algorithms. In~\cite{Khattak08,Mayer06}, the authors discuss strategies for base-station cooperation and study the effect of quantizing the exchanged values, still assuming perfect channel knowledge.

In this paper, we study cooperative receiver processing in an interference channel and formulate it as probabilistic inference in factor graphs. We state a probabilistic model that explicitly incorporates the ability of the receivers to exchange a certain type of information. To infer the information bits, we apply a recently proposed inference framework that combines BP and the mean-field (MF) approximation~\cite{Riegler2012}. We obtain a distributed iterative algorithm within which all receivers iteratively perform channel and noise precision estimation, detection and decoding, and also pass messages along the edges in the factor graph that connect them. The rate of updating and passing these messages determines the amount of communication over the cooperation links.

\emph{Notation}: The relative complement of $\{i\}$ in a set $\mathcal{I}$ is written as $\I\setminus i$. The set $\{ i \in \mathbb{N} \mid 1\leq i \leq n\}$ is denoted by $[1:n]$. Boldface lowercase and uppercase letters are used to represent vectors and matrices, respectively; superscripts ${(\cdot)}^{\operatorname{T}}$ and ${(\cdot)}^{\operatorname{H}}$ denote transposition and Hermitian transposition, respectively. The Hadamard product of two vectors is denoted by $\odot$.
The probability density function (pdf) of a multivariate complex Gaussian distribution with mean $\muv$ and covariance matrix $\Sgm$ is denoted by $\text{CN}(\cdot;\muv,\Sgm)$;
the pdf of a Gamma distribution with scale $a$ and rate $b$ is denoted by $\text{Ga}(\cdot;a,b)$.
We write $f(x) \propto g(x)$ when $f(x) = cg(x)$ for some positive constant $c$.
The Dirac delta function is denoted by $\delta(\cdot)$. Finally, $\text{E}[\cdot]$ stands for the expectation of a random variable.

\section{System Model} \label{SystModel}

We consider a system with $K$ parallel point-to-point links where each user sends information to its corresponding receiver and interferes with the others by doing so.
To decode the desired messages, the receivers are able to cooperate by exchanging information over dedicated error-free links.

A message sent by user $k$ is represented by a vector $\uv_k \in\{0,1\}^{I_k}$ of $I_k$ information bits and is conveyed by sending $N$ data and $L$ pilot channel symbols having the sets of indices $\D\subset{[1:N+L]}$ and $\P\subset[1:N+L]$, respectively, such that $\D\cup\P=[1:N+L]$ and $\D\cap\P=\emptyset$; the sets $\D$ and $\P$ are identical for all $K$ users.
The bits in $\uv_k$ are encoded and interleaved into a vector
$\cv_k \in \{0,1\}^{C_k}$ of $C_k=M_kN$ bits which are then mapped to data symbols $\xv_k^\text{D} = \left( x_k(i) \mid i\in \D \right)^{\operatorname{T}} \in \S_k^N$, where $\S_k$ is a (user specific) discrete complex modulation alphabet of size $2^{M_k}$.
Symbols $\xv_k^\text{D}$ are multiplexed with pilot symbols $\xv_k^\text{P} = \left( x_k(j) \mid j \in \P \right)^{\operatorname{T}}$ which are randomly drawn from a QPSK modulation alphabet.

The users synchronously transmit their aggregate vectors of channel symbols $\xv_k = \left( x_k(i) \mid i\in [1:N+L] \right)^{\operatorname{T}}$ over an interference channel with input-output relationship
\begin{equation}\label{eq:channel_IO}
\yv_l = \sum_{k\in[1:K]}\hv_{lk} \odot \xv_k + \wv_l, \quad \forall l\in[1:K].
\end{equation}
The vector $\yv_l = \left( y_l(i) \mid i\in [1:N+L] \right)^{\operatorname{T}}$ contains the signal received by receiver $l$, $\hv_{lk} = \left( h_{lk}(i) \mid i\in [1:N+L] \right)^{\operatorname{T}}$ is the vector of complex weights of the channel between transmitter $k$ and receiver $l$, and $\wv_l = \left( w_l(i) \mid i\in [1:N+L] \right)^{\operatorname{T}}$ contains the samples of additive noise at receiver $l$ with pdf $p(\wv_l) = \text{CN} \left( \wv_l;\mathbf{0},\gamma_l^{-1}\mathbf{I}_{N+L} \right)$ for some positive precision $\gamma_l$.
For all $l\in[1:K]$, we define the signal-to-noise ratio (SNR) and interference-to-noise ratio (INR) at receiver $l$ as
\begin{equation*} 
    \text{SNR}_l = \gamma_l \frac{\text{E}[\|\hv_{ll}\|^2]}{N+L}, \quad \text{INR}_l = \gamma_l \frac{\sum_{k\in[1:K] \setminus l} \text{E}[\|\hv_{lk}\|^2]}{N+L} .
\end{equation*}

\section{The Combined BP-MF Inference Framework} \label{BP-MF}

In this section, we consider a generic probabilistic model and briefly describe the unified message-passing algorithm that combines the BP and MF approaches~\cite{Riegler2012}.

Let $p(\zv)$ be an arbitrary pdf of a random vector
$\zv \triangleq \left( z_i \mid i \in \I \right)^{\operatorname{T}}$ which factorizes as
\begin{equation}\label{eq:arbit_pdf}
p(\zv) = \prod_{a \in \A}f_a(\zv_a) = \prod_{a\in
\A_{\text{MF}}}f_a(\zv_a) \prod_{c\in \A_{\text{BP}}}f_c(\zv_c)
\end{equation}
where $\zv_a$ is the vector of all variables $z_i$ that are arguments of the function $f_a$ for all $a \in \A$.
We have grouped the factors into two sets that partition $\A$: $\A_{\text{MF}}\cap \A_{\text{BP}}=\emptyset$ and $A_{\text{MF}}\cup \A_{\text{BP}}=\A$.
The factorization in~\eqref{eq:arbit_pdf} can be visualized by means of a factor graph~\cite{Kschischang2001}.
We define $\N(a)\subseteq\I$ to be the set of indices of all variables $z_i$ that are arguments of function $f_a$;
similarly, $\N(i)\subseteq\A$ denotes the set of indices of all functions $f_a$ that have variable $z_i$ as an argument.
The parts of the graph that correspond to $\prod_{a\in \A_{\text{BP}}}f_a(\zv_a)$ and to $\prod_{a\in
\A_{\text{MF}}}f_a(\zv_a)$ are referred to as ``BP part'' and ``MF
part'', respectively.

The combined BP-MF inference algorithm approximates the marginals
$p(z_i)=\int p(\zv)\prod_{j\in\I\setminus i} dz_j$, $i\in\I$, by auxiliary
pdfs $b_i(z_i)$ called \emph{beliefs}. They are computed as \cite{Riegler2012}
\begin{equation}\label{eq:beliefBPMF}
    b_i(z_i) = \omega_i \prod\limits_{c \in \A_\text{BP} \cap \N(i)} m^{\text{BP}}_{c\to i}(z_i)
                   \prod\limits_{c \in \A_\text{MF} \cap \N(i)} m^{\text{MF}}_{c\to i}(z_i),
\end{equation}
with
\begin{equation} \label{eq:updateBPMF}
\begin{split}
m^{\text{BP}}_{a\to i}(z_i) &= \omega_a
\int \prod_{j\in \N(a)\setminus i} \operatorname{d} \! z_j \,n_{j\to a}(z_j) \,f_a(\zv_a),\\
&\forall\ a\in \A_\text{BP}, i\in\N(a), \\
m^{\text{MF}}_{a\to i}(z_i) &=
\exp \left( \int \prod_{j\in \N(a)\setminus i} \operatorname{d} \! z_j \, n_{j\to a}(z_j) \ln f_a(\zv_a) \right),\\
&\forall\ a\in \A_\text{MF}, i\in\N(a), \\
n_{i\to a}(z_i) &= \omega_i \prod\limits_{c \in \A_\text{BP} \cap \N(i)
\setminus a} m^{\text{BP}}_{c\to i}(z_i) \prod\limits_{c \in \A_\text{MF} \cap \N(i)} m^{\text{MF}}_{c\to i}(z_i),\\
&\forall\ i\in\N(a), a\in\A,
\end{split}
\end{equation}
where $\omega_i$ and $\omega_a$ are constants that ensure normalized beliefs.

\section{Distributed Inference Algorithm} \label{Algorithm}
In this section, we state a probabilistic formulation of cooperative receiver processing and use the combined BP-MF framework to obtain the message updates in the corresponding factor graph; finally, we define a parametric iterative algorithm for distributed receiver processing.
\subsection{Probabilistic system model}

The probabilistic system function can be obtained by factorizing the joint pdf of all unknown variables in the signal model. Collecting the unknown variables in vector $\mathbf{v}$, we have:
\begin{equation} \label{eq:joint_pdf}
\begin{split}
p(\mathbf{v}) \propto &\prod_{l\in[1:K]} \Bigg[ p(\yv_l|\hv_{l1},\ldots,\hv_{lK},\xv_1^\text{D},\ldots,\xv_K^\text{D},\gamma_l) \, p(\gamma_l) \\
&\prod_{k\in[1:K]} p(\hv_{lk}) \Bigg] \prod_{k\in[1:K]} p(\xv_k^\text{D}|\cv_k) \, p(\cv_k|\uv_k) \, p(\uv_k).
\end{split}
\end{equation}

To include in the probabilistic model the ability of the different receivers to exchange information of a certain type, we define an augmented pdf.
Depending on the type of shared information, several cooperative strategies can be devised: the receivers could exchange their current local knowledge about the modulated data symbols $\xv_k^\text{D}$, or coded and interleaved bits $\cv_k$, or information bits $\uv_k$. We focus on the case in which the receivers share information on $\cv_k$\footnote{The other alternatives can be implemented with straightforward modifications to the model presented in this section.}.
To construct the augmented pdf for this cooperation scenario, we replace each vector variable $\xv_k$ and $\cv_k$ with $K$ ``alias" variables $\xv_{k,l}=\xv_k$ and $\cv_{k,l}=\cv_k$, $k,l\in[1:K]$, which are constrained to be equal to the corresponding original variable. Keeping in mind that receiver $l$ is interested in decoding message $\uv_l$, the factorization of the augmented pdf reads
\begin{equation} \label{eq:aug_joint_pdf}
\begin{split}
p(\mathbf{v}^\prime) &\propto \prod_{l\in[1:K]} \Bigg[ p(\yv_l|\hv_{l1},\ldots,\hv_{lK},\xv_{1,l}^\text{D},\ldots,\xv_{K,l}^\text{D},\gamma_l) \\
& \phantom{\propto} \, p(\gamma_l) \prod_{k\in[1:K]} \left( p(\hv_{lk}) \, p(\xv_{k,l}^\text{D}|\cv_{k,l}) \right) p(\cv_{l,l}|\uv_l) \\
& \phantom{\propto} \prod_{i\in[1:I_l]} p(u_l(i)) \prod_{k\in[1:K]\setminus l} p(\cv_{k,l}|\cv_{k,k}) \Bigg]
\end{split}
\end{equation}
where $\mathbf{v}^\prime$ denotes the vector of all unknown variables in \eqref{eq:aug_joint_pdf}, including the alias variables. Next, we denote, define and group in sets the factors in \eqref{eq:aug_joint_pdf}. For all $l\in[1:K]$, the factors
\begin{align*}
    &f_{\text{O}_l}(\hv_{l1},\ldots,\xv_{1,l}^\text{D},\ldots,\gamma_l) \triangleq p(\yv_l|\hv_{l1},\ldots,\xv_{1,l}^\text{D},\ldots,\gamma_l) \\
    &= \prod_{i\in\D\cup\P} \text{CN} \Bigg( y_l(i); \sum_{k\in[1:K]} h_{lk}(i) x_{k,l}(i),\gamma_l^{-1} \Bigg)
\end{align*}
incorporate the observation vector $\yv_l$ and they form the set $\A_\text{O}$; the factors $f_{\text{N}_l}(\gamma_l) \triangleq p(\gamma_l)$ are the prior pdfs of the parameters $\gamma_l$ and they form the set $\A_\text{N}$; the factors $f_{\text{H}_{lk}}(\hv_{lk}) \triangleq p(\hv_{lk}) = \text{CN} \left( \hv_{lk}; \hat{\hv}_{lk}^\text{p},\Sgm_{\hv_{lk}}^\text{p} \right)$, $k\in[1:K]$, represent the prior pdfs of the vectors $\hv_{lk}$ and they form the set $\A_\text{H}$; denoting by $\cv_{k,l}^i$ the subvector of $\cv_{k,l}$ containing the bits mapped on $x_{k,l}(i)$ and by $\M_k(\cdot)$ the mapping function, for all $k$, the factors
\begin{align*}
    f_{\text{M}_{k,l}} \left( \xv_{k,l}^\text{D},\cv_{k,l} \right) &\triangleq p \left( \xv_{k,l}^\text{D} | \cv_{k,l} \right)\\
    &= \prod_{i\in\D} \delta \left( x_{k,l}(i) - \M_k(\cv_{k,l}^i) \right)
\end{align*}
account for the modulation mapping and they form the set $\A_\text{M}$; the factors $f_{\text{C}_l}(\cv_{l,l},\uv_{l}) \triangleq p(\cv_{l,l}|\uv_l)$ stand for the coding and interleaving operations performed at transmitter $l$ and they form the set $\A_\text{C}$; the factors $f_{\text{U}_l^m}(u_l(m)) \triangleq p(u_l(m)), \, m\in[1:I_l]$ are the uniform prior probability mass functions of the information bits and they form the set $\A_\text{U}$; finally, for all $k \neq l$, the factors
\begin{equation} \label{eq:ExchgFactor}
\begin{split}
    f_{\text{E}_{kl}} (\cv_{k,l},\cv_{k,k}) &\triangleq p(\cv_{k,l}|\cv_{k,k}) \\
    &= \prod_{n\in[1:C_k]} \delta(c_{k,l}(n)-c_{k,k}(n))
\end{split}
\end{equation}
constrain the alias variables $\cv_{k,l}$, $l\in[1:K]$ to be equal, and they form the set $\A_\text{E}$. Note that, due to these additional constraints, marginalizing \eqref{eq:aug_joint_pdf} over all alias variables $\cv_{k,l}$, $l\neq k$ leads to the original probabilistic model \eqref{eq:joint_pdf}.

\begin{figure}[!t]
  \centering
  \includegraphics[width=\columnwidth]{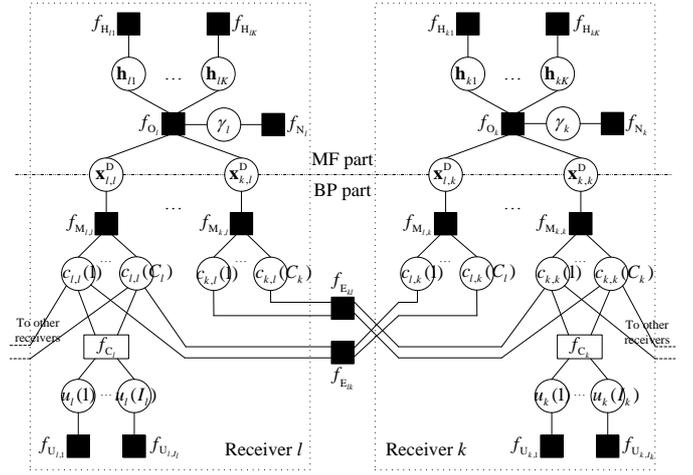}
  \caption{Factor graph representation of the pdf factorization in \eqref{eq:aug_joint_pdf}: receivers $l$ and $k$ are depicted together with the connections between them. For all $l\in[1:K]$, the bits $\cv_{l,l}$ in receiver $l$ are connected to the bits $\cv_{l,\cdot}$ in all other receivers, while the bits $\cv_{k,l}$, $k\neq l$, are only connected to the bits $\cv_{k,k}$ in receiver $k$.}
  \label{fig:FacGraph}
\end{figure}

The factorization in \eqref{eq:aug_joint_pdf} can be visualized in a factor graph, which is partially depicted in Fig.~\ref{fig:FacGraph}.
The graphs corresponding to the channel codes and interleavers are not given explicitly, their structures being captured by $f_{\text{C}_l}$.
We coin ``receiver $l$'' the subgraph containing the factor nodes $f_{\text{H}_{l1}},\ldots,$$f_{\text{H}_{lK}}$, $f_{\text{O}_{l}}$, $f_{\text{N}_{l}}$, $f_{\text{M}_{1,l}},\ldots,f_{\text{M}_{K,l}}$, $f_{\text{C}_{l}}$, $f_{\text{U}_{l,1}},\ldots,f_{\text{U}_{l,I_l}}$ and the variable nodes connected to them.
The factor nodes $f_{\text{E}_{lk}}$ and $f_{\text{E}_{kl}}$ model the cooperative link between receivers $l$ and $k$.

We can now recast the problem of cooperative receiver processing as an inference problem on the augmented probabilistic model~\eqref{eq:aug_joint_pdf}: receiver $l$ needs to infer the beliefs of the information bits in $\uv_l$ using the observation vector $\yv_l$ and prior knowledge, i.e., the pilot symbols of all users\footnote{Since the pseudo-random pilot sequences can be generated deterministically based on some information available to all receivers, each receiver is able to reconstruct all the pilot symbols without the need of exchanging them.} and their set of indices $\P$, the channel statistics, the modulation mappings of all users, the structure of the channel code and interleaver of user $l$, and the external information provided by the other receivers. The inference problem is solved by applying the method described in Section~\ref{BP-MF}, which leads to iteratively passing messages in the factor graph.
We can control the communication overhead between receivers by adjusting the rate of passing messages through nodes $f_{\text{E}_{lk}}$ and $f_{\text{E}_{kl}}$.

%

\subsection{Message computations}\label{Messages}

To make the connection with the arbitrary model in Section~\ref{BP-MF}, we define $\A$ and $\I$ to be the sets of all factors and variables, respectively, introduced in the previous subsection\footnote{With a slight abuse of notation, from this point on we use the names of functions and variables as indices in the sets $\A$ and $\I$, respectively.}.
We choose to split $\A$ into the following two sets that yield the ``MF part'' and the ``BP part'':
\begin{equation}\label{eq:split}
        \A_\text{MF} \triangleq \A_\text{H} \cup \A_\text{O} \cup \A_\text{N};\quad \A_\text{BP} \triangleq \A_\text{M} \cup \A_\text{C} \cup \A_\text{U} \cup \A_\text{E}.
\end{equation}

In the following, we use \eqref{eq:updateBPMF} to derive messages in our setup, focusing on their final expressions. More detailed message computations using the combined BP-MF method can be found in \cite{Riegler2012} and \cite{makirichfl11} for non-cooperative scenarios.

First, for all $k,l\in[1:K]$ we define the statistics
\begin{equation*}
    \hat x_{k,l}(i) \triangleq \sum_{\xv^\text{D}_{k,l}} n_{\xv^\text{D}_{k,l}\to f_{\text{O}_l}}(\xv^\text{D}_{k,l}) \, x_{k,l}(i),
\end{equation*}
\begin{equation*}
    \sigma^2_{x_{k,l}(i)} \triangleq \sum_{\xv^\text{D}_{k,l}} n_{\xv^\text{D}_{k,l}\to f_{\text{O}_l}}(\xv^\text{D}_{k,l}) \, |x_{k,l}(i)-\hat x_{k,l}(i)|^2
\end{equation*}
for $i\in\D$ and we set $\hat x_{k,l}(i) = x_{k,l}(i)$ and $\sigma^2_{x_{k,l}(i)} = 0$, for $i\in\P$.
We also define\footnote{The defined quantities are the parameters determining the corresponding beliefs, because $f_{\text{O}_l}$ is in the MF part and therefore the beliefs are equal to the ``$n$'' messages (see \eqref{eq:beliefBPMF},\eqref{eq:updateBPMF}).} $\hat \gamma_l \triangleq \int n_{\gamma_l\to f_{\text{O}_l}}(\gamma_l) \, \gamma_l \operatorname{d} \!\gamma_l$, $\hat \hv_{lk} \triangleq \int n_{\hv_{lk}\to f_{\text{O}_l}}(\hv_{lk}) \, \hv_{lk} \operatorname{d} \!\hv_{lk}$,
\begin{equation*}
    \Sgm_{\hv_{lk}} \triangleq \int n_{\hv_{lk}\to f_{\text{O}_l}}(\hv_{lk}) \, (\hv_{lk}-\hat\hv_{lk})(\hv_{lk}-\hat\hv_{lk})^{\operatorname{H}}  \operatorname{d} \!\hv_{lk},
\end{equation*}
and we denote by $\sigma^2_{h_{lk}(i)}$ the $(i,i)$th entry of $\Sgm_{\hv_{lk}}$.

\emph{Channel estimation}: Using \eqref{eq:updateBPMF}, we obtain the messages
\begin{equation} \label{eq:m_fO_h}
    \begin{split}
        m^{\text{MF}}_{f_{\text{O}_l} \to \hv_{lk}}(\hv_{lk}) & \propto \prod_{i\in\D\cup\P} \text{CN} \left( h_{lk}(i); \hat h^{\text{o}}_{lk}(i) , \sigma^2_{h^{\text{o}}_{lk}(i)} \right) \\
        & \propto \text{CN} \left( \hv_{lk}; \hat \hv^{\text{o}}_{lk} , \Sgm^{\text{o}}_{\hv_{lk}} \right), 
    \end{split}
\end{equation}
for all $k,l$, where $\Sgm^{\text{o}}_{\hv_{lk}}$ is a diagonal covariance matrix and
\begin{align*} 
    \hat h^{\text{o}}_{lk}(i) &= \frac{ \hat x_{k,l}^{\ast}(i) }{\sigma^2_{x_{k,l}(i)}+ |\hat x_{k,l}(i)|^2} \Bigg( y_l(i) - \sum_{k'\neq k} \hat{h}_{lk'}(i) \hat{x}_{k',l}(i) \Bigg) , \\
    \sigma^{-2}_{h^{\text{o}}_{lk}(i)} &= \hat \gamma_l \left( \sigma^2_{x_{k,l}(i)}+ |\hat x_{k,l}(i)|^2 \right), \quad \forall i \in \D \cup \P.
\end{align*}
We have $m^{\text{MF}}_{f_{\text{H}_{lk}} \to \hv_{lk}}(\hv_{lk}) = f_{\text{H}_{lk}}(\hv_{lk})$; so, using \eqref{eq:updateBPMF}, we obtain
\begin{equation} \label{eq:n_h_fO}
    n_{\hv_{lk}\to f_{\text{O}_l}}(\hv_{lk}) = \text{CN} \left( \hv_{lk}; \hat \hv_{lk} , \Sgm_{\hv_{lk}} \right), 
\end{equation}
$k,l\in[1:K]$, with
\begin{align*} 
    \Sgm_{\hv_{lk}}^{-1} &= \left(\Sgm^{\text{p}}_{\hv_{lk}}\right)^{-1} + \left(\Sgm^{\text{o}}_{\hv_{lk}}\right)^{-1}, \\
    \hat \hv_{lk} &= \Sgm_{\hv_{lk}} \left[ \left(\Sgm^{\text{p}}_{\hv_{lk}}\right)^{-1} \hat \hv^{\text{p}}_{lk} + \left(\Sgm^{\text{o}}_{\hv_{lk}}\right)^{-1} \hat \hv^{\text{o}}_{lk} \right].
\end{align*}

\emph{Noise precision estimation}: Using \eqref{eq:updateBPMF}, we obtain
\begin{equation} \label{eq:m_fO_g}
    m^{\text{MF}}_{f_{\text{O}_l} \to \gamma_l}(\gamma_l) \propto \gamma_l^{a_\text{o}} \exp(-d_\text{o}\gamma_l) \propto \text{Ga}(\gamma_l,a_\text{o}+1,d_\text{o}),
\end{equation}
$l\in[1:K]$, with $a_\text{o} = N+L$ and
\begin{align*} 
    d_\text{o} =& \sum_{i\in\D\cup\P} \Bigg[ \Big| y_l(i) - \sum_{k} \hat{h}_{lk}(i) \hat{x}_{k,l}(i) \Big|^2 + \sum_{k} \sigma^2_{x_{k,l}(i)}\sigma^2_{h_{lk}(i)}\\
    &+ \sum_{k} \sigma^2_{h_{lk}(i)}|\hat x_{k,l}(i)|^2 + \sum_{k} \sigma^2_{x_{k,l}(i)}|\hat h_{lk}(i)|^2  \Bigg].
\end{align*}
We select the conjugate prior pdf $f_{\text{N}_l}(\gamma_l) \triangleq \text{Ga}(\gamma_l,a_\text{p},d_\text{p})$, $l\in[1:K]$. Using \eqref{eq:updateBPMF}, we obtain
\begin{equation} \label{eq:n_g_fO}
    n_{\gamma_l \to f_{\text{O}_l}}(\gamma_l) = \text{Ga}(\gamma_l,a_\text{p}+a_\text{o},d_\text{p}+d_\text{o}).
\end{equation}
Setting the prior pdfs to be non-informative, i.e., $a_\text{p}=d_\text{p}=0$, we obtain the estimate $\hat \gamma_l = a_\text{o}/d_\text{o}$, $l\in[1:K]$.

\emph{Symbol detection}: Using \eqref{eq:updateBPMF}, we obtain
\begin{equation} \label{eq:m_fO_x}
    m^{\text{MF}}_{f_{\text{O}_l} \to \xv^\text{D}_{k,l}}(\xv^\text{D}_{k,l}) \propto \prod_{i\in\D} \text{CN} \left( x_{k,l}(i); \hat x^{\text{o}}_{k,l}(i) , \sigma^2_{x^{\text{o}}_{k,l}(i)} \right)
\end{equation}
with
\begin{align*} 
    \hat x^{\text{o}}_{k,l}(i) &= \frac{ \hat h_{lk}^{\ast}(i) }{\sigma^2_{h_{lk}(i)}+ |\hat h_{lk}(i)|^2} \Bigg(y_l(i) - \sum_{k'\neq k} \hat{h}_{lk'}(i) \hat{x}_{k',l}(i) \Bigg), \\
    \sigma^{-2}_{x^{\text{o}}_{k,l}(i)} &= \hat \gamma_l \left( \sigma^2_{h_{lk}(i)}+ |\hat h_{lk}(i)|^2 \right), \quad \forall i \in \D.
\end{align*}
Assume that in the BP part of the graph we have obtained 
\begin{equation} \label{eq:m_fM_x}
    m^{\text{BP}}_{f_{\text{M}_{k,l}} \to \xv^\text{D}_{k,l}}(\xv^\text{D}_{k,l}) \propto \prod_{i\in\D} \left( \sum_{s \in \S_k} \beta_{x_{k,l}(i)}(s)\delta(x_{k,l}(i)-s) \right),
\end{equation}
where $\beta_{x_{k,l}(i)}(s)$ is the extrinsic value of $x_{k,l}(i)$ for $s\in\S_k$.
According to \eqref{eq:updateBPMF}, the discrete messages (APP values)
\begin{align}
    &n_{\xv^\text{D}_{k,l} \to f_{\text{O}_l}}(\xv^\text{D}_{k,l}) \propto m^{\text{BP}}_{f_{\text{M}_{k,l}} \to \xv^\text{D}_{k,l}}(\xv^\text{D}_{k,l}) \, m^{\text{MF}}_{f_{\text{O}_l} \to \xv^\text{D}_{k,l}}(\xv^\text{D}_{k,l}) \label{eq:_n_x_fO}\\
    &\propto \prod_{i\in\D} \sum_{s \in \S_k} \beta_{x_{k,l}(i)}(s) \text{CN} \left( s; \hat x^{\text{o}}_{k,l}(i) , \sigma^2_{x^{\text{o}}_{k,l}(i)} \right) \delta(x_{k,l}(i)-s) \nonumber
\end{align}
are sent to the MF part, while $n_{\xv^\text{D}_{k,l} \to f_{\text{M}_{k,l}}}(\xv^\text{D}_{k,l}) \propto m^{\text{MF}}_{f_{\text{O}_l} \to \xv^\text{D}_{k,l}}(\xv^\text{D}_{k,l})$ are sent to the BP part as extrinsic values.

\emph{(De)mapping, decoding, information exchange}:  These operations are obtained using \eqref{eq:updateBPMF}, which due to \eqref{eq:split} reduce to the BP computation rules.
Messages from and to binary variable nodes are of the form $\theta\delta(v-0)+(1-\theta)\delta(v-1)$, with $\theta\in[0,1]$.
Computing $m^{\text{BP}}_{f_{\text{M}_{k,l}} \to c_{k,l}(n)}(c_{k,l}(n))$ is equivalent to MAP demapping, $k,l\in[1:K]$, $n\in[1:C_k]$.
The messages $n_{u_l(m) \to f_{\text{C}_l}}(u_l(m)) = f_{\text{U}_{l,m}}(u_l(m))$, $\forall m\in[1:I_l]$, and $n_{c_{l,l}(n) \to f_{\text{C}_l}}(c_{l,l}(n)) \propto m^{\text{BP}}_{f_{\text{M}_{l,l}} \to c_{l,l}(n)}(c_{l,l}(n))\times \prod_{k} m^{\text{BP}}_{f_{\text{E}_{lk}} \to c_{l,l}(n)}(c_{l,l}(n))$
represent the input values to the de-interleaving and decoding BP operations which output $m^{\text{BP}}_{f_{\text{C}_l} \to u_l(m)}$ and $m^{\text{BP}}_{f_{\text{C}_l} \to c_{l,l}(n)}$.
Due to the equality constraints \eqref{eq:ExchgFactor}, messages pass transparently through the factor nodes $f_{\text{E}_{lk}}$, $l,k\in[1:K]$.
Therefore, the following messages are received by receiver $k$ from receiver $l$, $k,l\in[1:K]$ and $n\in[1:C_k]$:
\begin{equation}
    m^{\text{BP}}_{f_{\text{E}_{kl}} \to c_{k,k}(n)}(c_{k,k}(n)) \propto  m^{\text{BP}}_{f_{\text{M}_{k,l}} \to c_{k,l}(n)}(c_{k,k}(n)),\label{eq:m_E_kk}
\end{equation}
\begin{equation}\label{eq:m_E_lk}
\begin{split}
    &m^{\text{BP}}_{f_{\text{E}_{lk}} \to c_{l,k}(n)}(c_{l,k}(n)) \propto m^{\text{BP}}_{f_{\text{M}_{l,l}} \to c_{l,l}(n)}(c_{l,k}(n)) \\
    &\times m^{\text{BP}}_{f_{\text{C}_l} \to c_{l,l}(n)}(c_{l,k}(n)) \times \prod_{k' \neq k} m^{\text{BP}}_{f_{\text{E}_{lk'}} \to c_{l,l}(n)}(c_{l,k}(n)).
\end{split}
\end{equation}
The messages 
\begin{align}
    n_{c_{k,l}(n) \to f_{\text{M}_{k,l}}}(c_{k,l}(n)) \propto &\, m^{\text{BP}}_{f_{\text{E}_{kl}} \to c_{k,l}(n)}(c_{k,l}(n)), \nonumber \\
    n_{c_{l,l}(n) \to f_{\text{M}_{l,l}}}(c_{l,l}(n)) \propto &\, m^{\text{BP}}_{f_{\text{C}_l} \to c_{l,l}(n)}(c_{l,l}(n)) \label{eq:n_c_fM}\\
    &\times \prod_{k} m^{\text{BP}}_{f_{\text{E}_{lk}} \to c_{l,l}(n)}(c_{l,l}(n)) \nonumber
\end{align}
$k,l\in[1:K]$, $n\in[1:C_k]$, are used in \eqref{eq:updateBPMF} to obtain the soft mapping updates \eqref{eq:m_fM_x}.

\subsection{Algorithm outline}
We define the cooperative processing algorithm by specifying the order in which the messages in Section~\ref{Messages} are computed and passed in the factor graph. The algorithm consists of three main stages: 
\subsubsection{Initialization}
Receiver $l$ obtains initial estimates of its variables.
First, estimates of $h_{kl}(i)$ with $i\in\P$ are obtained for all $k$ by using an iterative estimator based on the signals at pilot positions only, similar to the one described in \cite[Sec.\,V.A]{makirichfl11}. Specifically, we restrict \eqref{eq:m_fO_h}, \eqref{eq:n_h_fO}, \eqref{eq:m_fO_g} to include only subvectors and submatrices corresponding to pilot indices and we initialize $\hat \gamma_l=1$ and $\hat h_{kl}(i)=0$, $i\in\P$.
We compute \eqref{eq:m_fO_h} and \eqref{eq:n_h_fO} successively for all $k$, and then \eqref{eq:m_fO_g} and \eqref{eq:n_g_fO}; repeat this process $N_\text{in}$ times. The initial estimates of $\hv_{kl}$ are obtained by applying \eqref{eq:n_h_fO} for whole vectors and matrices, with $\hat h^{\text{o}}_{kl}(i)=\sigma^{-2}_{h^{\text{o}}_{lk}(i)}=0$, $i\in\D$. Then, we set $\hat x_{k,l}(i)=0$ and $\sigma^2_{x_{k,l}(i)}=1$, $i\in\D$.
Estimation of $\gamma_l$ is performed using \eqref{eq:m_fO_g} and \eqref{eq:n_g_fO}, followed by symbol detection \eqref{eq:m_fO_x}, applied successively for all $k$; this process is repeated $N_\text{det}$ times. Finally, soft demapping and decoding are performed in the BP part, with $m^{\text{BP}}_{f_{\text{E}_{kl}} \to c_{k,k}(n)}$ and $m^{\text{BP}}_{f_{\text{E}_{lk}} \to c_{l,k}(n)}$ initialized to have equal bit weights.
\subsubsection{Information exchange}
Receiver $l$ sends $m^{\text{BP}}_{f_{\text{E}_{kl}} \to c_{k,k}(n)}$ given by \eqref{eq:m_E_kk} to receiver $k$ and simultaneously receives $m^{\text{BP}}_{f_{\text{E}_{lk}} \to c_{l,l}(n)}$ from all receivers $k\neq l$; then, it computes and sends $m^{\text{BP}}_{f_{\text{E}_{lk}} \to c_{l,k}(n)}$ given by \eqref{eq:m_E_lk} to all receivers $k\neq l$.
\subsubsection{Local iteration}
Receiver $l$ computes \eqref{eq:n_c_fM}, followed by \eqref{eq:m_fM_x} and \eqref{eq:_n_x_fO}, for all $k$. Next, $\hv_{kl}$, $k\in[1:K]$, are successively estimated using \eqref{eq:m_fO_h} and \eqref{eq:n_h_fO}, and $\gamma_l$ is estimated using \eqref{eq:n_g_fO}. Then, \eqref{eq:m_fO_x} is successively computed for all $k$, repeating this process $N_\text{det}$ times. Finally, soft demapping and decoding are performed in the BP part.

To define the distributed iterative algorithm, we use three parameters: $N_\text{it}$ describes the total number of receiver iterations, including the \emph{Initialization} stage as first iteration; $N_\text{ex}\in[0:N_\text{it}-1]$ denotes the number of \emph{Information exchange} stages; for $N_\text{ex}>0$, the vector $\mathbf{t}_\text{E} = (t_\text{E}(e)\mid e \in [1:N_\text{ex}])\in [1:N_\text{it}-1]^{N_\text{ex}}$ with strictly increasing elements contains the iteration indices after which an \emph{Information exchange} stage takes place. For $N_\text{ex}=0$ we set $t_\text{E}=0$.
\begin{algorithm}\label{Algorithm}
The steps of the algorithm are:
\begin{enumerate}
\item \emph{Initialization} for all $l\in[1:K]$; Set $t=1$ and $e=1$;
\item If $t\geq N_\text{it}$ then go to step 5);
\item If $t_\text{E}(e)=t$ then \emph{Information exchange} $\forall l$; $e=e+1$;
\item \emph{Local iteration} for all $l$; $t=t+1$; go to step 2);
\item Take hard decisions using the beliefs \\ $b_{\text{U}_{l,m}}(u_l(m)) = \omega_{l,m}\, m^{\text{BP}}_{f_{\text{C}_l} \to u_l(m)}(u_l(m))\, f_{\text{U}_{l,m}}(u_l(m))$.
\end{enumerate}
\end{algorithm}

%
%

\section{Simulation Results}

\begin{figure}[!t]
\centering
\includegraphics[width=0.8\columnwidth]{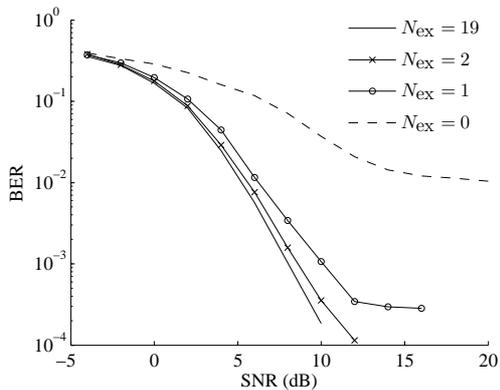}
\caption{BER vs. SNR performance of the distributed iterative algorithm.}
\label{fig:BER_SNR}
\end{figure}

\begin{figure}[!t]
\centering
\includegraphics[width=0.8\columnwidth]{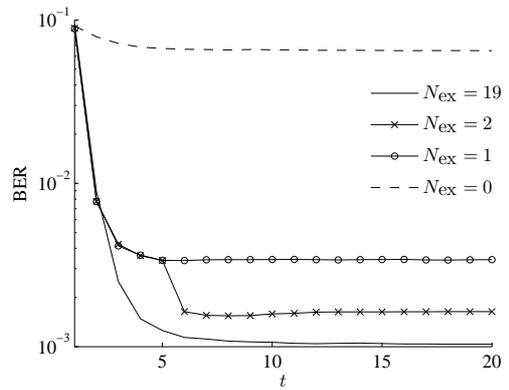}
\caption{BER vs. iteration number at $\text{SNR}=8$ dB.}
\label{fig:BER_iter}
\end{figure}

We consider an OFDM system consisting of $K=2$ links with symmetric channel powers, same noise levels at the receivers, and strong interference, i.e. $\text{SNR}_1=\text{SNR}_2=\text{INR}_1=\text{INR}_2=\text{SNR}$.
The detailed assumptions are listed in Table~\ref{tab:sim_param}.
The performance of Algorithm \ref{Algorithm} is evaluated through Monte-Carlo simulations.
The BER dependence on SNR is illustrated in Fig.~\ref{fig:BER_SNR}, while the BER convergence is given in Fig.~\ref{fig:BER_iter}.
Receiver collaboration provides a significantly improved performance compared to a non-cooperative setting ($N_\text{ex}=0$). When $N_\text{ex}=1$, an error-floor occurs at BER $\approx 3\cdot 10^{-4}$, but the cooperation scheme with only two exchanges almost achieves the performance of ``full'' cooperation ($N_\text{ex}=19$); the improvement brought by the second exchange is clearly visible in Fig.~\ref{fig:BER_iter}.
All schemes need about $5$--$6$ receiver iterations to converge.
The benefits of cooperation are also observed in the improved channel weights and noise precision estimation (results are not presented here), which of course lead to improved detection and decoding, and vice-versa.
\begin{table}[!h]
\caption{Simulation parameters}
\label{tab:sim_param}
\begin{tabular}{l|l}
\hline
\hline
\textbf{Parameters of the OFDM system} & \textbf{Value} \\
\hline
Number of users & $K=2$ \\
Subcarrier spacing & $15\,\text{kHz}$ \\
Number of active subcarriers & $N+L = 100$ \\
Number of pilot symbols & $L = 17$ evenly spaced pilots\\
Modulation scheme for data symbols & $\text{QPSK} \, (M_1=M_2=2)$\\
Convolutional code (of both users) & $R = 1/3,\,(133,171,165)_8$ \\
Multipath channel model & 3GPP ETU \\
\hline
\hline
\textbf{Parameters of the algorithm} & \textbf{Value} \\
\hline
Number of receiver iterations & $N_\text{it}=20$ \\
Number of exchanges & $N_\text{ex} \in \{0, 1, 2, 19 \}$  \\
Exchange indices & $\mathbf{t}_\text{E} \in \{0, 1, (1,5),(1,\ldots,19)\}$  \\
Number of sub-iterations & $N_\text{in}=10$, $N_\text{det}=5$ \\
\hline
\end{tabular}
\end{table}
\section{Conclusions}
We proposed a message-passing design of a distributed algorithm for receiver cooperation in interference-limited wireless systems. Capitalizing on a unified inference method that combines BP and the MF approximation,
we obtained an iterative algorithm that jointly performs estimation of channel weights and noise powers, detection, decoding in each receiver and information sharing between receivers. Simulation results showed a remarkable improvement compared to a non-cooperative system, even with 1--2 exchanges between receivers; as expected, a trade-off between performance and amount of shared information could be observed.

In general, our approach provides several degrees of freedom in the design of distributed algorithms, such as the type of shared information and the parameters of the algorithm (number of receiver iterations, rate and schedule of information exchange). The proposed approach can be extended to other cooperation setups and it can accommodate the exchange of quantized values -- the quantization resolution thus becoming another implementation choice -- by quantizing the parameters of the messages passed between the receivers.






%
\bibliographystyle{IEEEtran}
\bibliography{IEEEabrv,References}

\end{document}